\documentclass{INTERSPEECH2023}
\usepackage{amsmath,graphicx}
\usepackage{color}
\usepackage{amssymb}
\usepackage{booktabs}
\usepackage{hyperref}
\usepackage{multirow}
\usepackage{comment}

\interspeechcameraready

\title{Improving Generalization Ability of Countermeasures for New Mismatch Scenario by Combining Multiple Advanced Regularization Terms}
\name{Chang Zeng$^{1,2}$, Xin Wang$^1$, Xiaoxiao Miao$^1$, Erica Cooper$^1$, Junichi Yamagishi$^{1,2}$}
\address{
  $^1$National Institute of Informatics, Japan $^2$SOKENDAI, Japan}
\email{\{zengchang, wangxin, xiaoxiaomiao, ecooper, jyamagis\}@nii.ac.jp}

\begin{document}

\maketitle
 
\begin{abstract}
\vspace{-2mm}
The ability of countermeasure models to generalize from seen speech synthesis methods to unseen ones has been investigated in the ASVspoof challenge. However, a new mismatch scenario in which fake audio may be generated from real audio with unseen genres has not been studied thoroughly. To this end, we first use five different vocoders to create a new dataset called CN-Spoof based on the CN-Celeb1\&2 datasets. Then, we design two auxiliary objectives for regularization via meta-optimization and a genre alignment module, respectively, and combine them with the main anti-spoofing objective using learnable weights for multiple loss terms. The results on our cross-genre evaluation dataset for anti-spoofing show that the proposed method significantly improved the generalization ability of the countermeasures compared with the baseline system in the genre mismatch scenario.
\end{abstract}


\vspace{-2mm}
\section{Introduction}
\vspace{-1mm}

With the development of deep neural networks, DeepFake detection has attracted much attention from academia and industry since synthesized media, such as video and speech by artificial intelligence, has brought huge risks. The ASVspoof challenge series \cite{asvspoof2015,asvspoof2017,asvspoof2019,asvspoof2021} has been proposed to explore and promote the investigation of anti-spoofing for automatic speaker verification (ASV) systems. The Audio Deep Synthesis Detection (ADD) challenge \cite{add2022} was held to study countermeasure models in a scenario with low-quality audio. The generalization ability in anti-spoofing is an important topic, and typical distribution shifts between training and testing data, such as unseen synthesis methods \cite{asvspoof2019} and unseen acoustic environments \cite{asvspoof2021}, have been investigated in the challenges. However, there is room for further investigations.

\begin{figure}[t]
    \centering
    \includegraphics[width=7.5cm]{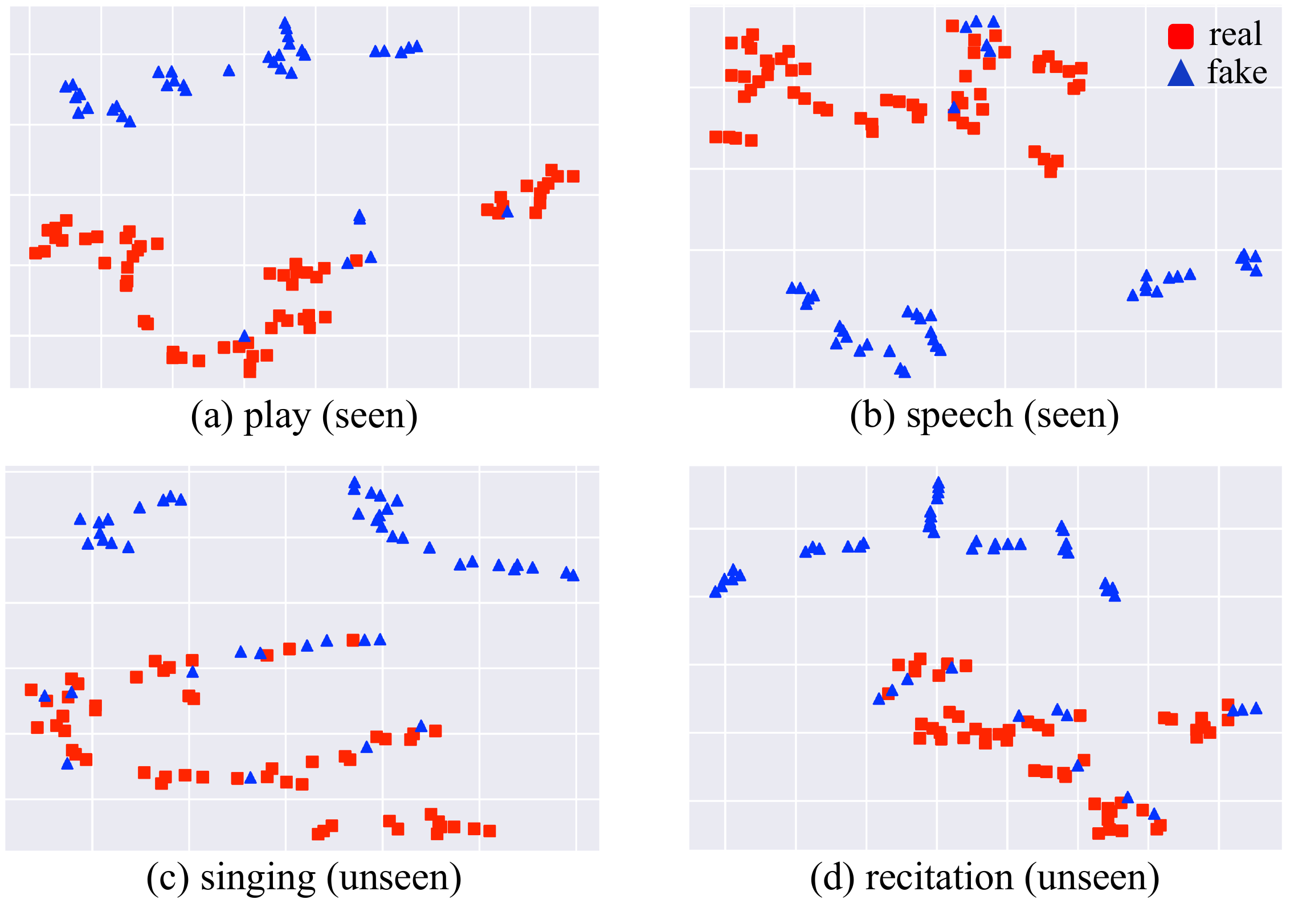}
    \caption{Visualization of LCNN countermeasure embedding by T-SNE.}
    \label{fig:lcnn-tsne}
\vspace{-6mm}
\end{figure}

Here, we focus on a new mismatch scenario related to the audio genre, in which fake audio may be generated from real audio with unseen genres because the effects of the speaker's styles and intrinsic factors associated with the specific genre have not been studied thoroughly in anti-spoofing. The definition of audio genre in our study is the same as \cite{cn1,cn2}. 
To analyze the effects and propose a more robust system, we utilize the copy-synthesis method \cite{copy-synthesis, copy-synthesis2, wavefake, spoofed2022} to produce spoofed data using multiple vocoders from real mel-spectrograms extracted from the waveforms of the CN-Celeb1\&2 datasets \cite{cn1,cn2}, which contain more than $600,000$ utterances with $11$ different genres. The new dataset is called CN-Spoof. To visually show the genre mismatch, we randomly select eight genres from the CN-Celeb1\&2 and CN-Spoof datasets and train a lightweight convolutional neural network (LCNN) \cite{asvspoof2019, lcnn}. We use the well-trained LCNN to extract the countermeasure embeddings from the waveforms with two seen and two unseen genres and visualize them in two-dimensional space by T-SNE \cite{tsne} as Fig.\ \ref{fig:lcnn-tsne} shows. From the figure, it is obvious that the embedding with seen ``play'' and ``speech'' genres can be classified well by the LCNN model. However, for the embedding with unseen ``singing'' and ``recitation'' genres, part of the real and fake embeddings overlap, which means it is hard to distinguish the fake audio samples with unseen genres by the LCNN model.

To address this genre mismatch, we make four training protocols by grouping different genres. For each protocol, we randomly select utterances from several genres as the training dataset, and all training protocols share the same evaluation dataset. 
On the basis of the training and evaluation datasets, in this paper, we propose a novel multi-task learning method and design two auxiliary regularization objectives in addition to the main anti-spoofing objective to improve the generalization ability of the countermeasure model. The first objective is to simulate the genre mismatch scenario in the training stage by meta-optimization \cite{maml,robustmaml,domain-invariant,meta-cross-channel}. In addition, for the second objective, we utilize a genre alignment module that contains a gradient reversal layer (GRL) \cite{grl,grl-spk} to remove the genre information in the countermeasure embedding. Finally, the two auxiliary regularization objectives are combined with the main anti-spoofing objective by using uncertainty loss weights \cite{mtl1,mtl2}, which can be learned from the data instead of setting them by hand resulting in inferior optimization. The experimental results on the evaluation dataset show that the proposed method significantly improves the genre generalization ability compared with the baseline system.

The rest of this paper is organized as follows. In Section \ref{sec:dataset}, we describe how to generate the CN-Spoof dataset and protocols for studying genre mismatch. The proposed multi-task learning method is illustrated in Section \ref{sec:method}. The experimental results are shown in Section \ref{sec:exp}. Finally, we conclude the paper in Section \ref{sec:con}.

\begin{figure}[t]
    \centering
    \includegraphics[width=8cm]{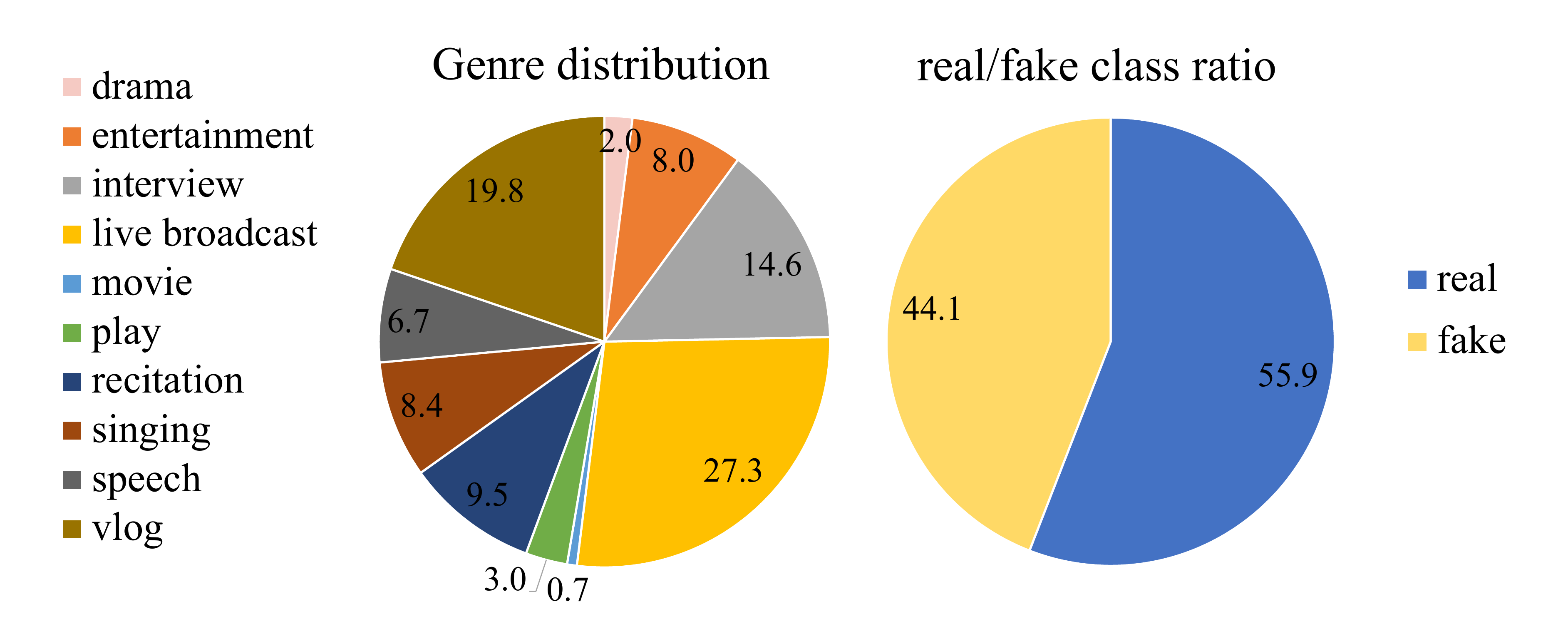}
    \caption{Genre distribution and real/fake class ratio for the evaluation dataset.}
    \label{fig:eval_data}
\vspace{-2mm}
\end{figure}

\vspace{-2mm}
\section{CN-Spoof dataset and protocols}
\vspace{-1mm}
\label{sec:dataset}
Since there are no related datasets in the research on anti-spoofing for the genre mismatch scenario, we leverage three pretrained neural vocoders (Multi-band MelGAN \cite{mbmelgan}, Parallel WaveGAN \cite{pwg}, and HiFiGAN \cite{hifigan}), and two DSP vocoders (WORLD \cite{world} and Griffin-Lim \cite{griffin-lim}), to reconstruct the waveforms from the real mel-spectrograms of the CN-Celeb1\&2 datasets which are collected for speaker verification in the multi-genre scenario. Specifically, for each vocoder, we randomly select $20,000$ and $80,000$ utterances from the CN-Celeb1 and CN-Celeb2 datasets, respectively. Then, vocoded waveforms of the selected utterances using the above five vocoders are treated as fake data. As a result, the CN-Spoof dataset contains $500,000$ fake utterances in total.

After generating the CN-Spoof dataset, we combine it with the CN-Celeb1\&2 datasets and randomly sample $200,000$ utterances from all genres as the evaluation dataset, whose genre distribution and real/fake class ratio are shown in Fig.\ \ref{fig:eval_data}. The remaining portions are utilized to construct the training dataset for different cross-genre protocols (CGP). We first divide ten genres included in the remaining portions into four groups, shown in Table \ref{tab:genre-group}. Note that the ``advertisement" genre is discarded from the dataset since its number is $2,929$, which is limited. For each CGP, we randomly sample $660,000$ utterances with three genre groups as the training dataset from the remaining portions. In this way, each training dataset has an unseen genre group that exists in the evaluation dataset, as Table \ref{tab:protocol} shows.

\begin{table}[t]
  \caption{Genre group division}
  \label{tab:genre-group}
  \centering
  \begin{tabular}{lc}
    \toprule
    \multicolumn{1}{l}{\textbf{Group}} & \multicolumn{1}{c}{\textbf{Genre Types}} \\
    \midrule
    Group I & drama (dr), vlog (vl), speech (sp) \\
    \midrule
    Group II & entertainment (en), interview (in), play (pl) \\
    \midrule
    Group III & live broadcast (lb), movie (mo) \\
    \midrule
    Group IV & singing (si), recitation (re) \\
    \bottomrule
  \end{tabular}
\vspace{-4mm}
\end{table}

\begin{table}[t]
  \caption{Cross-genre protocols (CGP)}
  \setlength\tabcolsep{4pt}
  \label{tab:protocol}
  \centering
  \begin{tabular}{lccc}
    \toprule
    \multicolumn{1}{l}{\textbf{CGP}} & \multicolumn{1}{c}{\textbf{Seen Genres}} & \multicolumn{1}{c}{\textbf{Unseen Genres}} \\
    \midrule
    CGP I & Group I, Group II, Group III & Group IV \\
    \midrule
    CGP II & Group I, Group II, Group IV & Group III \\
    \midrule 
    CGP III & Group I, Group III, Group IV & Group II \\
    \midrule
    CGP IV & Group II, Group III, Group IV & Group I \\
    \bottomrule
  \end{tabular}
\vspace{-4mm}
\end{table}

\begin{figure*}[t]
    \centering
    \includegraphics[width=17cm]{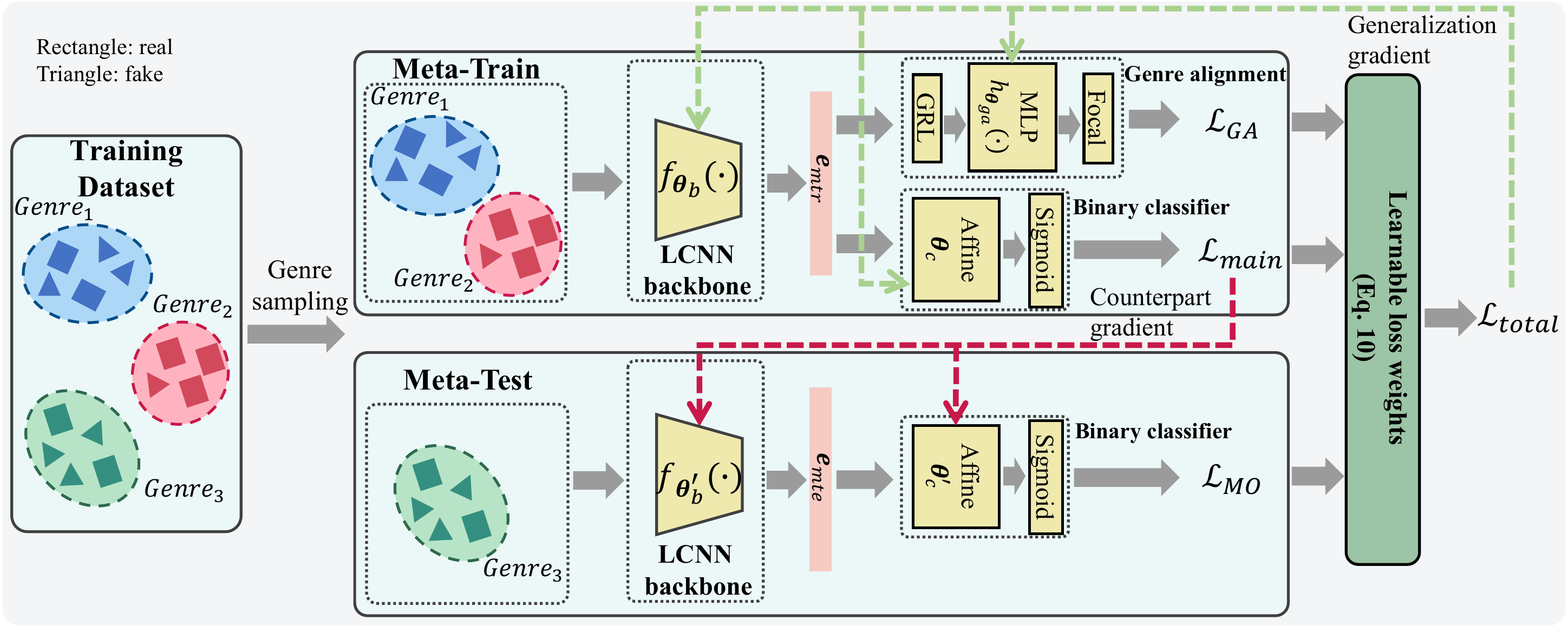}
    \caption{Architecture of proposed multi-task learning method.}
    \label{fig:arch}
\vspace{-4mm}
\end{figure*}

\vspace{-2mm}
\section{Proposed multi-task learning method}
\vspace{-1mm}
\label{sec:method}
The proposed multi-task learning method is illustrated concretely in this section. Since our method contains meta-optimization and genre alignment regularization objectives, we first describe a genre sampling strategy for constructing a task set for all objectives. Then, the main objective and auxiliary regularization objectives are depicted, respectively. Finally, we give details on combining multiple objectives with the uncertainty loss weights.

\subsection{Genre sampling}
\vspace{-2mm}
As shown in Fig.\ \ref{fig:arch}, we randomly sample data from the training dataset $\mathcal{D} = \{\mathcal{D}_1, ..., \mathcal{D}_G|G > 1\}$, which includes $G$ seen genres, to construct a task set $\mathcal{T} = \{\mathcal{T}_1, ..., \mathcal{T}_K|K > 1\}$, which contains $K$ tasks. A task is composed of a meta-train dataset that contains $G_{mtr} (G_{mtr} < G)$ genres and is used for the main anti-spoofing objective as well as the auxiliary genre alignment objective, and a meta-test dataset is used that contains the remaining $G_{mte} = G - G_{mtr}$ genres for the auxiliary meta-optimization objective. In our experiments, for each training protocol, we set $G_{mtr}$ as $G - 1$ for the meta-train dataset, and the remaining one is used as the meta-test dataset.

\subsection{Main anti-spoofing objective}
\vspace{-2mm}
The main anti-spoofing objective is to distinguish whether input speech is real. In our proposed method, LCNN \cite{asvspoof2019,lcnn} is selected as the backbone, as shown in Fig.\ \ref{fig:arch}. Countermeasure embeddings are extracted by the LCNN model from the meta-train dataset, which is represented by a formula:
\begin{align}
    \boldsymbol{e}_{mtr} = f_{\boldsymbol{\theta}_b}(\boldsymbol{x}_{mtr}),
\end{align}
where $\boldsymbol{e}_{mtr}$ and $\boldsymbol{x}_{mtr}$ represent the countermeasure embedding and corresponding speech in the meta-train dataset, respectively. $f_{\boldsymbol{\theta}_b}(\cdot)$ means the transformation function of the LCNN backbone, whose parameters are $\boldsymbol{\theta}_b$.

Then, a binary classifier, including an affine layer and a Sigmoid function, transforms the embeddings $\boldsymbol{e}_{mtr}$ into probability values, representing the possibility of speech being real. Finally, we perform a binary cross-entropy (BCE) loss function on the probability values and the corresponding labels. The process can be formulated as:
\begin{align}
    P(\boldsymbol{e}_{mtr}) & = \frac{1}{1+\exp(-\boldsymbol{e}_{mtr}^\top\boldsymbol{\theta}_c)}, \\
    \mathcal{L}_{main} & = -\frac{1}{N}\sum_{i = 1}^{N}[\mathcal{I}(y^{i}_{mtr} = 1)\log P(\boldsymbol{e}^{i}_{mtr}) \nonumber \\
    & + \mathcal{I}(y^{i}_{mtr} \neq 1)\log (1 - P(\boldsymbol{e}^{i}_{mtr}))], 
    \label{eq:main}
\end{align}
where $\mathcal{L}_{main}$ represents the cost of the main anti-spoofing objective on the meta-train dataset, which has $N$ speech samples. $\boldsymbol{\theta}_c$ denotes the parameters of the classifier. Note that for concise description, we ignore the bias parameter of the affine layer in Eq. (2). $\mathcal{I}(\cdot)$ is an indicator function that returns $1$ if the condition is true and $0$ otherwise, and $y^i_{mtr}$ is the $i$-th label. When $\boldsymbol{e}^{i}_{mtr}$ is from real speech, $y^i_{mtr}$ equals $1$, otherwise $0$. In the following part, we use $\boldsymbol{\theta}$ to denote the $\boldsymbol{\theta}_b$ and $\boldsymbol{\theta}_c$.

\subsection{Meta-optimization regularization objective}
\vspace{-2mm}
In the genre sampling stage, we divide a task into a meta-train and a meta-test dataset without overlapping genres. In this way, we can simulate the genre mismatch in the training stage by evaluating the performance on the meta-test dataset of the model trained on the meta-train dataset. Specifically, as Fig.\ \ref{fig:arch} shows, we maintain a counterpart of the model parameters and update it by the loss $\mathcal{L}_{main}$ with the learning rate $\beta$:
\begin{align}
\label{eq:counterpart}
    \boldsymbol{\theta}' = \boldsymbol{\theta} - \beta * \frac{\partial \mathcal{L}_{main}}{\partial \boldsymbol{\theta}},
\end{align}
where $\boldsymbol{\theta}'$ denotes the updated parameters of the counterpart, including $\boldsymbol{\theta}'_b$ and $\boldsymbol{\theta}'_c$. Next, we evaluate the model performance on the meta-test dataset and compute the loss:
\begin{align}
    \boldsymbol{e}_{mte} & = f_{\boldsymbol{\theta}'_b}(\boldsymbol{x}_{mte}), \\
    P(\boldsymbol{e}_{mte}) & = \frac{1}{1+\exp(-\boldsymbol{e}_{mte}^\top\boldsymbol{\theta}'_c)}, \nonumber \\
    \mathcal{L}_{MO} & = -\frac{1}{M}\sum_{i = 1}^{M}[\mathcal{I}(y^{i}_{mte} = 1)\log P(\boldsymbol{e}^{i}_{mte}) \\
    & + \mathcal{I}(y^{i}_{mte} \neq 1)\log (1 - P(\boldsymbol{e}^{i}_{mte}))],
\end{align}
where $\boldsymbol{x}^{i}_{mte}$ and $y^{i}_{mte}$ are the $i$-th speech sample and corresponding label in the meta-test dataset, which has $M$ samples. $\boldsymbol{e}_{mte}$ is the corresponding countermeasure embedding. $\mathcal{L}_{MO}$ denotes the loss on the meta-test dataset for the meta-optimization regularization objective.

\subsection{Genre alignment regularization objective}
\vspace{-2mm}
In addition to simulating the genre mismatch in the training stage by meta-optimization, we use another auxiliary genre alignment regularization objective that is realized by using a multi-layer perceptron (MLP) with the GRL component to improve the generalization ability of the LCNN model because it can remove the genre information contained in countermeasure embeddings by adversarial training \cite{grl}. Before inputting the countermeasure embedding $\boldsymbol{e}_{mtr}$ to the MLP for genre classification, we first apply the GRL component to it as Fig.\ \ref{fig:arch} shows. The MLP contains three layers, and each has $128$ neural units. Instead of utilizing plain cross-entropy as the loss function, here we use focal loss \cite{focal} because it can improve the discriminative ability of the genre alignment module, which is beneficial for further filtering out the genre information in countermeasure embeddings. The loss function of the genre alignment module is formulated as:
\begin{align}
    P(\boldsymbol{e}_{mtr}, g_c) & = \frac{\exp(h_{\boldsymbol{\theta}_{ga}^{g_c}}(\boldsymbol{e}_{mtr}))}{\sum_{j=1}^{G}\exp(h_{\boldsymbol{\theta}_{ga}^{g_j}}(\boldsymbol{e}_{mtr}))}, \label{eq:softmax} \\
    \mathcal{L}_{GA} & = -\frac{1}{N}\sum_{i}^{N}(1-P(\boldsymbol{e}_{mtr}^i, g_{c}^i))^\gamma \log P(\boldsymbol{e}_{mtr}^i, g_{c}^i), \label{eq:focal}
\end{align}
where $g_c$ is the genre label of the countermeasure embedding $\boldsymbol{e}_{mtr}$. $h_{\boldsymbol{\theta}_{ga}^{g_c}}(\cdot)$ denotes a transformation for outputting the un-normalized possibility of the $c$-th genre from the genre alignment module, whose parameters are $\boldsymbol{\theta}_{ga}$. $P(\boldsymbol{e}_{mtr}, g_c)$ is the probability that the countermeasure embedding $\boldsymbol{e}_{mtr}$ belongs to the correct genre, and $\gamma$ is a hyper-parameter that can adjust the gradient contribution of different samples in accordance with their difficulties. Note here that we ignore the hyper-parameter $\alpha$ in \cite{grl} since all genres are treated equally in our method.
Due to the GRL component, the gradient after the GRL in the backward propagation is reversed compared with that without the GRL.

\begin{table*}[t]
  \caption{EER (\%) of experimental results on CGP. For each protocol, the genre group in the bracket does not appear in the training dataset. A bold number means the best performance of this genre.}
  \label{tab:result}
  \setlength\tabcolsep{5pt}
  \centering
  \begin{tabular}{ccccccccccccc} 
    \toprule
    \multirow{2}{*}{\textbf{Protocol}} & \multirow{2}{*}{\textbf{System}} & \multirow{2}{*}{\textbf{Overall}} & \multicolumn{3}{c}{\textbf{Group I}} & \multicolumn{3}{c}{\textbf{Group II}} & \multicolumn{2}{c}{\textbf{Group III}} & \multicolumn{2}{c}{\textbf{Group IV}} \\
    \cmidrule(l{0em}r{0em}){4-13}
    &  &  & \textbf{dr} & \textbf{vl} & \textbf{sp} & \textbf{en} & \textbf{in} & \textbf{pl} & \textbf{lb} & \textbf{mo} & \textbf{si} & \textbf{re} \\
    \toprule
    \multirow{3}{*}{\textbf{CGP I}} & $\mathcal{L}_{main}$ & 8.299 & 6.890 & 9.124 & 6.582 & 7.505 & 7.799 & 6.876 & 7.933 & 7.960 & 9.517 & 9.779 \\
    \multirow{3}{*}{(Group IV)} & $\mathcal{L}_{MO}, \mathcal{L}_{main}$ & 7.863 & 5.626 & 9.053 & 6.565 & 5.962 & 6.818 & 6.148 & 7.929 & 6.799 & 8.761 & 9.385 \\
    & $\mathcal{L}_{GA}, \mathcal{L}_{main}$ & 8.238 & 6.831 & 9.082 & 6.904 & 7.399 & 7.672 & 6.855 & 7.903 & 8.031 & 9.063 & 9.615  \\
    & $\mathcal{L}_{MO}, \mathcal{L}_{GA}, \mathcal{L}_{main}$ & \textbf{7.511} & \textbf{5.109} & \textbf{8.827} & \textbf{6.362} & \textbf{5.508} & \textbf{6.266} & \textbf{5.216} & \textbf{7.577} & \textbf{6.799} & \textbf{8.248} & \textbf{9.157} \\
    \toprule 
    \multirow{3}{*}{\textbf{CGP II}} & $\mathcal{L}_{main}$ & 8.566 & 7.176 & 9.281 & 6.887 & 7.601 & 8.147 & 7.414 & 8.919 & 7.794 & 9.053 & 8.996 \\
    \multirow{3}{*}{(Group III)} & $\mathcal{L}_{MO}, \mathcal{L}_{main}$ & 8.181 & 5.926 & 9.387 & 6.797 & 6.456 & 7.184 & 6.408 & 8.682 & 6.965 & 8.916 & 8.788 \\
    & $\mathcal{L}_{GA}, \mathcal{L}_{main}$ & 8.481 & 7.176 & 9.369 & 7.073 & 7.320 & 7.863 & 6.855 & 8.787 & 7.334 & 9.157 & 9.110 \\
    & $\mathcal{L}_{MO}, \mathcal{L}_{GA}, \mathcal{L}_{main}$ & \textbf{7.764} & \textbf{5.788} & \textbf{8.959} & \textbf{6.664} & \textbf{5.770} & \textbf{6.676} & \textbf{5.365} & \textbf{8.339} & \textbf{6.347} & \textbf{8.314} & \textbf{8.425} \\
    \toprule
    \multirow{3}{*}{\textbf{CGP III}} & $\mathcal{L}_{main}$ & 8.599 & 7.922 & 9.099 & 7.035 & 8.620 & 8.983 & 8.905 & 8.112 & 8.679 & 9.424 & 8.603 \\
    \multirow{3}{*}{(Group II)} & $\mathcal{L}_{MO}, \mathcal{L}_{main}$ & 8.182 & 7.118 & 8.942 & 6.823 & 7.657 & 8.236 & 7.489 & \textbf{7.641} & \textbf{7.772} & 9.277 & 8.693 \\
    & $\mathcal{L}_{GA}, \mathcal{L}_{main}$ & 8.505 & 8.266 & 8.915 & 6.785 & 8.365 & 9.009 & 8.420 & 7.947 & 8.808 & 9.658 & 8.657 \\
    & $\mathcal{L}_{MO}, \mathcal{L}_{GA}, \mathcal{L}_{main}$ & \textbf{8.032} & \textbf{6.889} & \textbf{8.746} & \textbf{6.447} & \textbf{7.464} & \textbf{8.192} & \textbf{7.202} & 7.738 & 8.126 & \textbf{8.983} & \textbf{8.060} \\
    \toprule
    \multirow{3}{*}{\textbf{CGP IV}} & $\mathcal{L}_{main}$ & 8.160 & 6.339 & 9.322 & 7.242 & 6.995 & 7.639 & 7.261 & 7.886 & 8.290 & 8.966 & 8.657 \\
    \multirow{3}{*}{(Group I)} & $\mathcal{L}_{MO}, \mathcal{L}_{main}$ & 7.827 & \textbf{5.454} & 9.357 & 7.174 & 6.141 & 6.765 & \textbf{5.602} & 7.850 & 6.965 & 8.502 & 8.790 \\
    & $\mathcal{L}_{GA}, \mathcal{L}_{main}$ & 7.944 & 6.028 & 9.334 & 7.124 & 6.954 & 7.131 & 6.483 & \textbf{7.819} & 6.799 & 8.804 & \textbf{8.299} \\
    & $\mathcal{L}_{MO}, \mathcal{L}_{GA}, \mathcal{L}_{main}$ & \textbf{7.739} & 5.568 & \textbf{9.182} & \textbf{7.073} & \textbf{6.004} & \textbf{6.489} & 5.700 & 7.840 & \textbf{6.136} & \textbf{8.341} & 8.633 \\
    \bottomrule
  \end{tabular}
\vspace{-3mm}
\end{table*}

\subsection{Learnable loss weights}
\vspace{-2mm}
\label{sec:llw}
As the proposed method has multiple loss terms, the model performance is extremely sensitive to loss weight selection \cite{mtl1,mtl2}.
To this end, we use a common strategy in multi-task learning to combine multiple loss terms with learnable loss weights, the aim of which is learning the optimal weights. Because our auxiliary objectives can be discarded in the inference stage, we treat the loss of auxiliary objectives as the regularization terms \cite{mtl2}. Thus, the total loss can be formulated as:
\begin{align}
\label{eq:total}
    \mathcal{L}_{total} = & \frac{1}{2*\lambda_{main}^2}*\mathcal{L}_{main} + \ln (1 + \lambda_{main}^2) \nonumber \\
    & + \frac{1}{2*\lambda_{MO}^2}*\mathcal{L}_{MO} + \ln (1 + \lambda_{MO}^2) \nonumber \\ 
    & + \frac{1}{2*\lambda_{GA}^2}*\mathcal{L}_{GA} + \ln (1 + \lambda_{GA}^2),
\end{align}
where $\lambda_{main}$, $\lambda_{MO}$, and $\lambda_{GA}$ are the learnable parameters for each loss term, respectively. As for the logarithmic terms in Eq. (\ref{eq:total}), they are constraint conditions for avoiding trivial solutions \cite{mtl1,mtl2}.

\section{Experimental results}
\vspace{-1mm}
\label{sec:exp}

\subsection{Experimental setup}
\vspace{-2mm}
As described in Section \ref{sec:dataset}, the CN-Spoof dataset was combined with the CN-Celeb1\&2 datasets, and we divided these data in accordance with the cross-genre protocols. For each protocol, the training dataset contained $660,000$ utterances. All protocols shared the same evaluation dataset, which contained $200,000$ utterances from all genres. 

As for the systems used for comparison in the paper, the LCNN model was selected as the baseline system. Additionally, we also constructed two other experimental systems: one incorporating the $\mathcal{L}_{main}$ and $\mathcal{L}_{MO}$ losses, and the other incorporating the $\mathcal{L}_{main}$ and $\mathcal{L}_{GA}$ losses. These systems were developed to thoroughly examine the impact of various regularizations. 

\vspace{-1mm}
\subsection{Training methodology}
\vspace{-2mm}
For the systems without meta-optimization, we randomly selected $64$ samples from the dataset as a mini-batch. For the systems including meta-optimization, we randomly selected $64$ samples from the dataset. One genre was randomly selected from this subset as the meta-test dataset, while the remaining samples were used as the meta-train dataset. We trained all systems for $40$ epochs using an SGD optimizer \cite{sgd} with a $0.001$ initial learning rate, $0.9$ momentum, and $0.0001$ L2 regularization. The learning rate was decayed by $0.9$ every epoch. As for other hyper-parameters, $\beta$ in Eq. (\ref{eq:counterpart}) was set to $0.001$, and $\gamma$ in Eq. (\ref{eq:focal}) was set to $5$.
\vspace{-1mm}
\subsection{Results and analysis}
\vspace{-2mm}
The experimental results are shown in Table \ref{tab:result}. As we can see, for each protocol, the countermeasure performance without the regularization terms on the unseen genres was generally worse than the counterparts for the other protocols except for the ``movie'' genre in CGP II and ``drama'' genre in CGP IV. This result proves that a model without the regularization terms cannot generalize well on unseen genres, which is consistent with our hypothesis.

Although the GRL component has shown the capacity to generalize well on unseen domains in the speaker verification task \cite{grl-spk}, we found that the genre alignment loss $\mathcal{L}_{GA}$ only slightly improved the generalization ability compared with the baseline model in the genre mismatch scenario. In contrast, the meta-optimization loss $\mathcal{L}_{MO}$ improved the EER numbers on some unseen genres, such as the ``singing'' genre for CGP I and ``movie'' genre for CGP II. However, there are some other genres that the system had difficulty generalizing well, such as the ``speech'' genre for CGP III. 

As for our multi-task learning method that integrates the meta-optimization and genre alignment regularization objectives using learnable loss weights, it significantly improved the generalization ability in the genre mismatch scenario by comparing its result with the baseline system for unseen genres. Even for some challenging genres like the ``recitation’' and ``speech’' genres, which cannot be generalized well by using either $\mathcal{L}_{MO}$ or $\mathcal{L}_{GA}$ only, combining them further improve the generalization ability.

Moreover, we can see that our proposed system not only performs well on unseen genres but also on seen genres. For protocol CGP I and II, our system obtained the best performance on all genres compared with the other systems. As for the protocol CGP III and IV, although our approach performed slightly worse than the other systems for several genres, it still obtained the best EER numbers for most genres. 

\section{Conclusions}
\vspace{-1mm}
\label{sec:con}
In this paper, we explored a new mismatch scenario for the anti-spoofing objective, in which the fake speech may come from the real speech with unseen genres. Since there is no anti-spoofing data related to this scenario, we utilized the copy-synthesis method to create a spoofed dataset called CN-Spoof based on the CN-Celeb1\&2 datasets. Our proposed multi-task learning method, which combines the meta-optimization loss and genre alignment loss as the regularization terms by using learnable loss weights, shows the potential to improve the generalization ability of the countermeasure models under this scenario. The experimental results on four different cross-genre protocols proved that our method is more robust than the baseline system, even when facing difficult genres. 

\section{Acknowledgements}
This study is partially supported by JST CREST Grants (JPMJCR18A6 and JPMJCR20D3), and MEXT KAKENHI Grants (21K17775, 21H04906, 21K11951, 22K21319).


\bibliographystyle{IEEEtran}
\bibliography{mybib}

\end{document}